\documentclass[11pt,a4paper]{article}
\usepackage{jinstpub}
\usepackage{rotating}
\usepackage{epstopdf}
\usepackage{epsfig}
\usepackage{graphicx}
\usepackage{upgreek}
\usepackage{caption}
\usepackage{subcaption}
\usepackage{array}
\usepackage{multirow}
\usepackage[utf8]{inputenc}

\begin{document}


\title{Characterization of Chromium Compensated GaAs as an x-ray Sensor Material for Charge-Integrating Pixel Array Detectors}

\author[a,b,1]{Julian Becker\note{now at Deutsches Elektronen-Synchrotron DESY, 22607 Hamburg, Germany},}
\author[a]{Mark W. Tate,}
\author[a]{Katherine S. Shanks,}
\author[a]{Hugh T. Philipp,}
\author[a,b]{Joel T. Weiss,}
\author[a]{Prafull Purohit,}
\author[b,\dag]{Darol Chamberlain\note[\dag]{in memoriam},}
\author[a,b,c,2]{and Sol M. Gruner\note{Corresponding author}}

\affiliation[a]{Laboratory of Atomic and Solid State Physics, Cornell University, Ithaca, NY 14853,USA}
\affiliation[b]{Cornell High Energy Synchrotron Source (CHESS), Cornell University, Ithaca, NY 14853, USA}
\affiliation[c]{Kavli Institute at Cornell for Nanoscale Science, Cornell University, Ithaca, NY 14853, USA}
\emailAdd{smg26@cornell.edu}

\keywords{Hybrid pixel detector, integrating detector, Hi-Z}

\abstract{We studied the properties of chromium compensated GaAs when coupled to charge
integrating ASICs as a function of detector temperature, applied bias and x-ray tube energy. The
material is a photoresistor and can be biased to collect either electrons or holes by the pixel circuitry.
Both are studied here. Previous studies have shown substantial hole trapping. This trapping and
other sensor properties give rise to several non-ideal effects which include an extended point spread
function, variations in the effective pixel size, and rate dependent offset shifts. The magnitude of
these effects varies with temperature and bias, mandating good temperature uniformity in the sensor
and very good temperature stabilization, as well as a carefully selected bias voltage.}

\maketitle

\section{Introduction}

Since the introduction of Pixel Array Detectors (PADs) in the late 1990s \cite{I1}, the detector landscape
at synchrotron sources has arguably changed. One of the advantages of PADs is the separation of the
sensor and signal processing into two distinct layers, keeping the Application Specific Integrated
Circuit (ASIC) apart from the sensitive volume, that detects radiation. In this way the sensor
material and readout can be optimized independently. The most commonly used sensor material to
date is silicon, typically with thicknesses in the range from 300~$\upmu$m to 500~$\upmu$m. Silicon sensors are
low cost, readily available and usually of excellent quality, especially when compared to alternative
sensor materials. Unfortunately, silicon x-ray sensors are less useful at higher x-ray energies because
the stopping power of silicon diminishes rapidly for x-rays above approximately 20 keV.


Existing detector solutions for high energy photons each have their own merits and flaws \cite{detectors}. In the energy band from approximately 15~--~50~keV, gallium arsenide (GaAs) is a potentially suitable sensor material candidate, and was studied previously by our group \cite{d1}.  

\begin{figure*}[tb]
  \centering
  \includegraphics[width=0.75\textwidth]{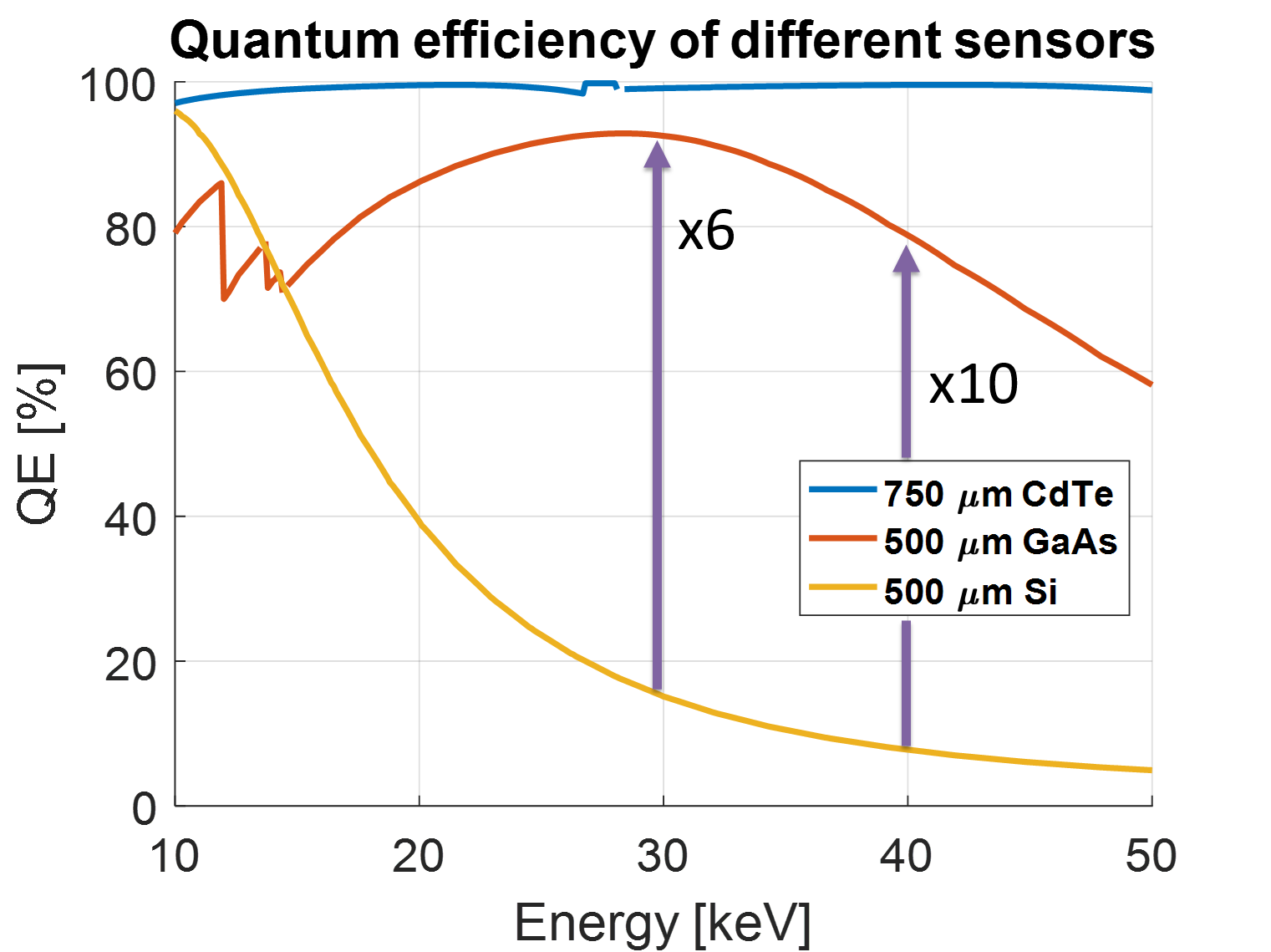}
  \caption{Quantum efficiency of 750~$\upmu$m CdTe and 500~$\upmu$m GaAs and silicon sensors. For energies above approximately 15~keV the probability to absorb a photon is consistently higher in GaAs than in silicon sensors; however it is also consistently lower than in CdTe sensors. The investigated GaAs sensors were produced with a 1~$\upmu$m gold layer on the entrance window, significantly reducing the sensitivity to low energy photons.}
  \label{QE}
\end{figure*}

For a sensor thickness of 500~$\upmu$m\footnote{A sensor thickness of 500~$\upmu$m is common for GaAs sensors.}, the quantum efficiency is high, as shown in figure \ref{QE}. Multiple studies using chromium compensated GaAs sensors (GaAs:Cr) bonded to the photon counting Medipix ASICs have reported good properties \cite{G3, G4, G5, G6, G7, G8, G9}. In the same energy range CdTe, another promising sensor material, has a similar detection efficiency. However CdTe shows strong effects of polarization and auto-fluorescence above approximately 30~keV \cite{static, dynamic}.

The auto-fluorescence of GaAs happens above approximately 11~keV, however since the mean distance to absorption is less than 50~$\upmu$m \cite{david} the effect is much smaller than that observed in CdTe and negligible in our case.

In this report we examine the properties of GaAs:Cr when coupled to charge integrating ASICs \cite{lpd}, as opposed to photon counting mode ASICs \cite{pil1, pil2, eiger, mp3}. The motivation is that photon counting PADs are unfeasible for use in situations where more than one x-ray arrives per pixel in the photon processing time, such as is commonly seen in x-ray free electron laser experiments and in many high instantaneous flux x-ray storage ring experiments. In these cases, integrating detectors are the only available option. 

\section{GaAs:Cr sensor material}

The chromium compensated GaAs material used in this study was produced by Tomsk State University (TSU) in Russia \cite{G1, G2}. It was used to fabricate sensor devices with 128~$\times$~128 pixels at a pitch of 150~$\upmu$m in both directions. The edge of the sensor is surrounded by a guard ring. The available sensors have an entry window electrode that consists of 1~$\upmu$m Au and 0.1~$\upmu$m Ni, which significantly reduces the sensitivity to x-rays below approximately 10~keV. TSU is currently developing a process to replace the Au layer with a less absorbent Al layer for future sensor fabrications.

The nominal average resistivity of the wafer used in this study was 0.5~G$\Omega$cm. Measurements were done using a contactless measurement technique \cite{G11, G12} at room temperature. 

Since the sensors are fabricated as photoresistors, the dark current is comparatively high and depends strongly on bias and temperature \cite{G10}. While this is not much of a problem for most photon counting detectors like the Medipix3 \cite{mp3}, it is an important parameter for integrating detectors, like the one used in this study.

The mobility of electrons and holes and their lifetimes was measured at TSU to be 2500 cm$^2$/Vs and 40~ns for electrons and 165 cm$^2$/Vs and 1.1 ns for holes \cite{G12,G10}. 

Given the sensor thickness of 500~$\upmu$m and a bias voltage of 200~V, we expect more than 88 \% of electrons to reach the readout electrode before being trapped\footnote{At the given mobility this translates to a charge collection time of 5~ns or less.}, however the probability of holes reaching the electrode before being trapped is practically zero. Even though the small pixel effect \cite{spe} works in our favor if we collect electrons, we still expect a noticeable effect from the trapping and detrapping of holes. 

Discussing the theory behind these expected effects is beyond the scope of this paper. However, we draw the attention of the reader to excellent review articles on this topic \cite{review, spe}, which explain the background and origin of the effects seen in our sensor material, but do not investigate GaAs material.

\section{The MM-PAD ASIC and detector system}


The MM-PAD ASIC \cite{M1, M2, M3, M4, mm_gain, M5} was developed as a wide dynamic range integrating detector system with single x-ray sensitivity. It was designed for hole collection and features a charge removal scheme that removes quantized amounts of charge from the integration node during the integration time and increments a digital counter for each removal process, thus increasing the total dynamic range.

To take advantage of the favorable electron transport properties of the material, most of the tests presented in this paper were performed with the chip operating in electron collection mode. The MM-PAD ASIC, however, was designed for hole collection and the charge removal process does not work properly with electron collection, hence it was disabled for the tests with electron collection\footnote{Making an MM-PAD ASIC designed to collect electrons in straightforward, but would have required a modification of the circuitry and a new fabrication run and  was not possible with the available budget.}. Therefore the dynamic range in most tests in this report is limited to the amount of charge that can be stored on the integration node. While this very small dynamic range limited the ability to collect images with high contrast, it still allowed much information to be gathered about imaging with GaAs:Cr sensors in a charge integrating mode, as shown herein. This information is very applicable to future wide dynamic range, electron collecting charge integrating detectors that use GaAs:Cr. 

For reasons that will be motivated by the experimental results, the preferred detector working point of our system is 15~C and -200~V bias (electron collection). We found this point was best to balance all the positive and negative effects of temperature and voltage. Please note than many results presented by others, especially those obtained with photon counters from the Medipix family are obtained at room temperature \cite{G5, G6, G7, G8, G9}.

\begin{figure*}[tb!]
  \centering
	\includegraphics[width=\textwidth]{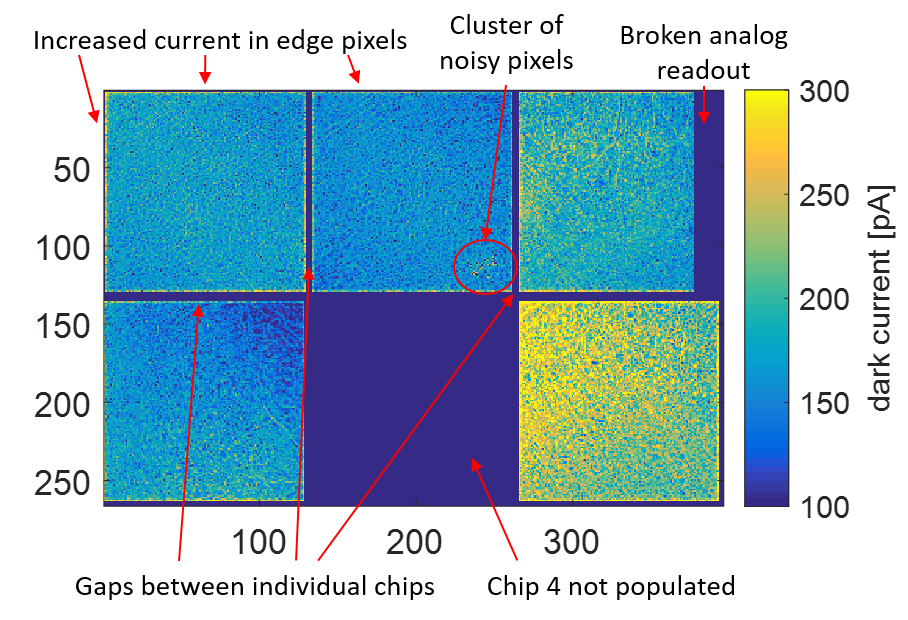}
  
  \caption{Per pixel dark current determined from integration time scans of dark images at a bias of -200~V at a temperature of 15~C. Several features of the system have been highlighted for reference. Chips are numbered starting in the upper left corner at chip 1 and descending along columns until reaching chip 6 in the lower right corner.}
  \label{dark}
\end{figure*}

Bump deposition and flip-chip bonding of GaAs:Cr sensors to MM-PAD ASICs were performed by Micross AIT \cite{micross}. Placement and gluing of the resultant modules onto heat sinks was performed in-house. Wire-bonding of the ASIC was done by Majelac Technologies LLC \cite{majelac}. 

In this paper the results obtained from 5 bump bonded ASICs (assemblies) mounted in a U-shape are presented. All of the studied sensors were manufactured on a single wafer. Figure \ref{dark} shows the arrangement of the sensors and certain prominent features, which are visible in the dark current map. We observe 3 dead or unbonded pixels in chip 2, a cluster of noisy pixels in chip 3 and a dead analog bank (16 pixel columns) in chip 5. We suspect the systematically higher dark current of chip 6 to be a result of insufficient thermal contact. In order to facilitate assembling and dis-assembling of the system the prototype investigated here was assembled without applying thermal grease to the interfaces.

\subsection{Data evaluation and correction}

This section is used to discuss the signals produced by the detector and the most common corrections applied to them in order to facilitate understanding of the experimental results.

As outlined in \cite{M3} the ASIC produces a digital number of the charge removal operations and
an analog remainder that is subsequently digitized. The detector electronics samples both signals
and combines them to a value in Analog-to-Digital converter Units (ADU). In electron collection
mode the digital number is always zero because charge removal is disabled.

These combined `raw' ADU values already provide useful information, but also have practical limitations. The most obvious limitation is that a `zero' reading does not produce an ADU number equal to zero. The reason for this is that all the elements in the readout chain have working points selected such that they do not operate at the extreme limits of their range.

Commonly we employ an offset correction to account for all effects causing a non-zero reading in absence of signal. The offset correction is determined by averaging a large number of dark frames, usually around 100. Dark frames are frames in absence of signal, e.g., frames taken with a closed x-ray shutter. This average dark frame is then subtracted from the signal frame to correct for the offsets of each individual pixel. Inspecting the offset correction for pixel-to-pixel variations and/or changes with bias and temperature provides insight into the mechanisms contributing to the offsets. In order for this correction to be valid, the time between determination of the offsets and the experiment should be small compared to the timescale of potential drift of these values. 

\begin{figure*}[tb]
  \centering
  \includegraphics[width=0.8\textwidth]{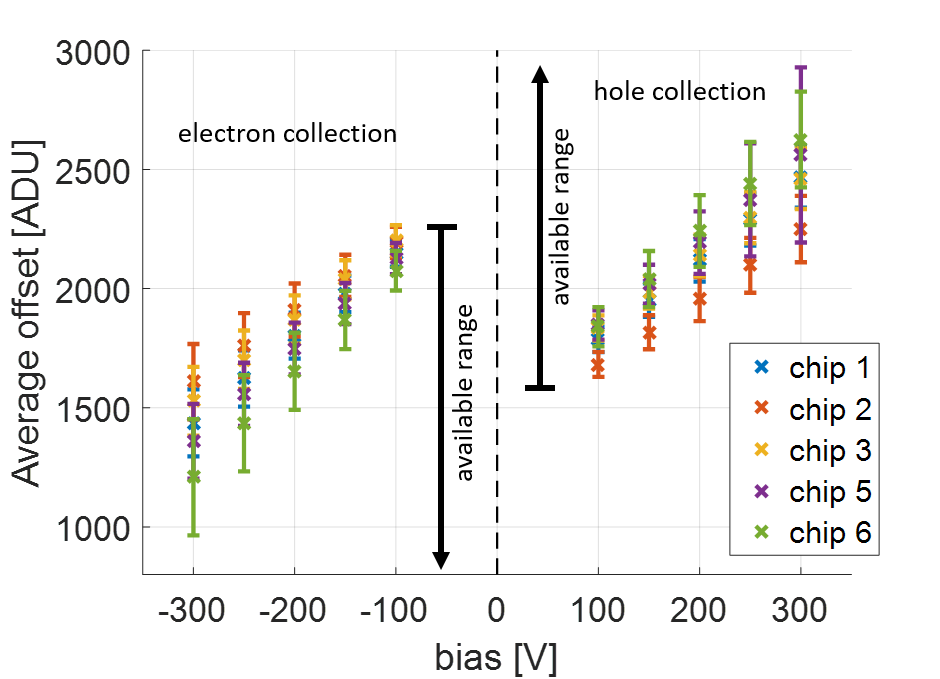}
  \caption{Average offset of the individual chips for 100~$\upmu$s integration time at 15~C as a function of bias. The available range of the ADC spans its full range from 0 to 4095. }
  \label{bkg}
\end{figure*}

After offset correction, the readings in absence of a signal will be distributed around zero, i.e.,
have a vanishing mean. The spread of the measurements arises from noise of the system, e.g., from
the readout electronics and the shot noise of the dark current. The mean magnitude of readout noise\footnote{The contribution of the readout was estimated by measuring the total noise as a function of integration time and extrapolating to zero integration time.} is about 1.2 keV, which is small compared to the signal from a single
x-ray in the 10~--~50 keV range. See below for details on the total noise as a function of various
parameters.

The MM-PAD ASIC has an additional common-mode offset of a few ADU that is global to the ASIC, but the level of which varies randomly with each read. The magnitude of the common-mode offset is comparable to the magnitude of the readout noise. In normal operation with Si sensors, this common-mode is sampled and compensated, a process called ‘debouncing’. Each MM-PAD ASIC has an offset value that is uncorrelated with the offsets of the other chips in the same detector system.

Since the common-mode offset is comparatively small, it is negligible in the presence of high flux illumination, and for low flux illumination can be corrected using statistical methods\footnote{The histogram of an individual image will show a strong noise peak with a center that is shifted from the zero position. The magnitude of this shift is used as a `debouncing' correction for this frame.} .

Note that the statistical nature of the common-mode offset causes the average of the per frame de-bounce corrections to vanish in the limit of an infinite number of frames. The experiments below will show situations where the de-bounce correction does not vanish, indicating the presence of an additional offset that is not corrected by the described offset subtraction.

Figure \ref{bkg} shows the average offset correction for a single frame as a function of applied bias. Positive biases and negative biases use different settings for the on-chip reference sources causing a discontinuity between the two polarities. 

Since charge removal does not work in electron collection mode, it is important that the offset value is as close to the upper limit\footnote{Electrons produce a negative signal.} of the Analog-to-Digital Converter (ADC) range as possible\footnote{The detector used in this investigation used 12-bit ADCs, making the maximum sample value $2^{12}-1=4095$.} to maximize the available dynamic range of the signal\footnote{The dynamic range of the signal can be limited by other elements, e.g. buffers, along the readout chain as well. This is not the case in our system.}. We note that due to the resistive nature of the sensor the dark current increases with increased bias voltage, hence the dynamic range available to record the signal is reduced at higher bias voltages. This observation is in contrast to the common experience with silicon sensors that are built as photodiodes. In silicon sensors the dark current increases until the sensor is fully depleted and varies very little with further increases of the bias voltage until the diode breaks down.

\section{Characterization under equilibrium conditions}
As shown in the next section, it is important to note if results are obtained under equilibrium
conditions or in a temporally transient state of the material, such as after changes in temperature, bias or illumination. All results presented in this section were obtained under equilibrium conditions with 100~$\upmu$s integration time. At 15~C this allowed a dynamic range of several hundreds of ADU, even in high current pixels, before clipping of the signal occurred.

The time required to reach equilibrium depends on the temperature of the sensor, as shown in
the next section. It took approximately 1 hour after each temperature step for the sensor to come
to thermal equilibrium. This is much longer that in silicon and is, most likely due to the missing
thermal grease on the thermal interfaces and the fact that the thermal conductivity of GaAs being
approximately 3 times lower than that of silicon. Illumination response equilibrium was achieved
by illuminating the sensor for at least 10 seconds before taking data.

We used two different x-ray sources in this section. An Am-241 source was used to determine the gain of the system and look at single event distributions. All other measurements were performed using a 50~W silver (Ag) anode x-ray tube (Trufocus TCM-5000M).

\subsection{Gain}

\begin{figure*}[tb]
  \centering
  \begin{subfigure}[t]{0.45\textwidth}
	\includegraphics[width=\textwidth]{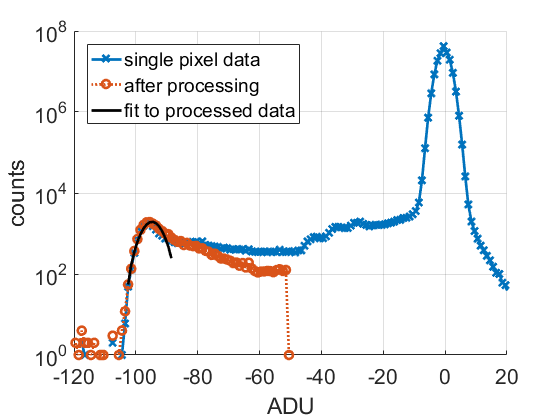}
	\caption{Histogram for -200~V applied bias (electron collection).}
	\label{hist_a}
  \end{subfigure}
\quad
  \begin{subfigure}[t]{0.45\textwidth}
	\includegraphics[width=\textwidth]{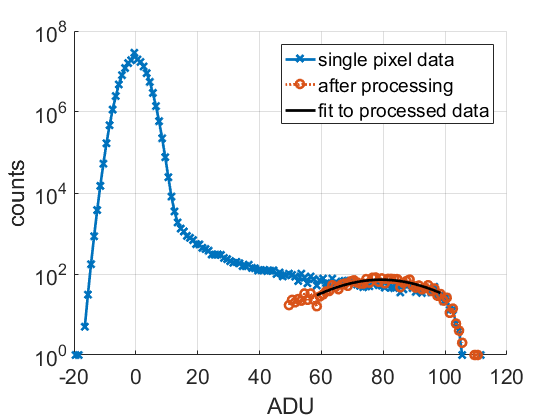}
	\caption{Histogram for +200~V applied bias (hole collection).}
	\label{hist_b}
  \end{subfigure}
  
  \caption{Histograms of the data collected at 15~C with the Am-241 source. Single pixel data were offset corrected and de-bounced (blue), processed (red) and the photopeak was fit using the processed data (black).}
  \label{histograms}
\end{figure*}

We used the 59.5~keV line of Am-241 to calibrate the detector response as a function of temperature and applied bias.

Before evaluation the data were offset corrected and de-bounced, and split events were summed together using the algorithms developed previously in our lab during the investigations of CdTe sensor material \cite{static}. While split events\footnote{We assume that most of the observed split events are due to a fraction of the charge cloud extending to the neighboring pixel(s), i.e., charge sharing, not due to fluorescence events.} commonly only extended to one neighbor pixel in electron collection mode, we had to sum over a larger cluster of pixels in order to reliably measure the deposited signal in hole collection mode. The reason for this will become clear when the results of the edge spread function test are presented (see below). The resulting photon histograms for electron and hole collection at -200~V and +200~V at a temperature of 15~C are shown in figure \ref{histograms}.

\begin{figure*}[tb]
  \centering
  \includegraphics[width=0.8\textwidth]{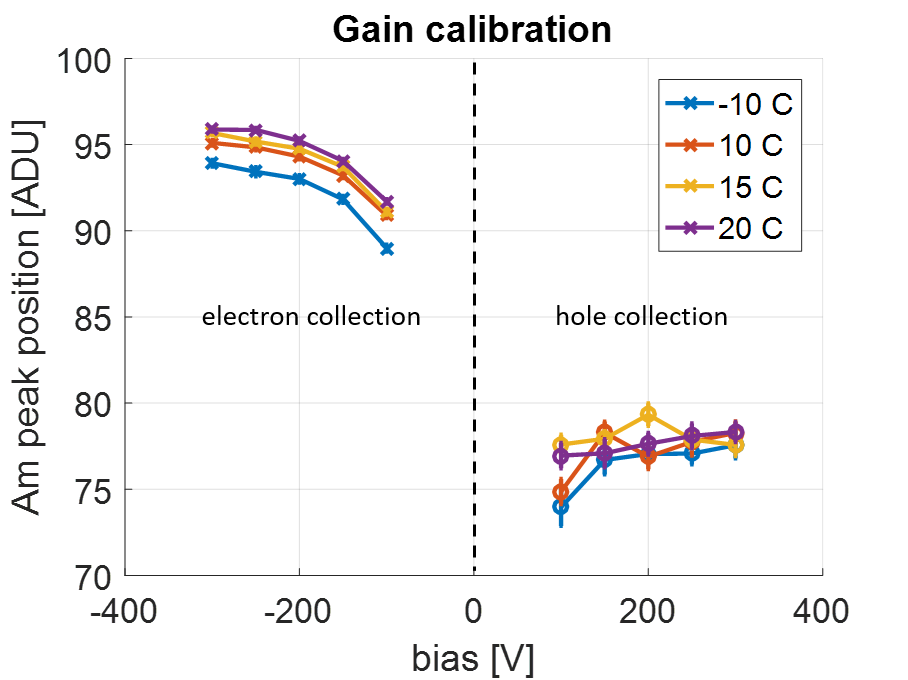}
  \caption{Position of the Am-241 photopeak (59.5~keV) as a function of temperature and bias. The peak position corresponds to the average of the total energy deposited in a single cluster, which can span several pixels in the case of hole collection.}
  \label{gain}
\end{figure*}

\begin{figure*}[tb!]
  \centering
  \begin{subfigure}[t]{0.45\textwidth}
	\includegraphics[width=\textwidth]{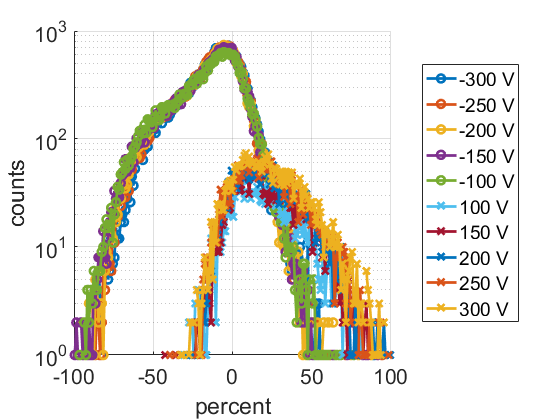}
	\caption{Data including split events.}
	\label{halo_a}
  \end{subfigure}
\quad
  \begin{subfigure}[t]{0.45\textwidth}
	\includegraphics[width=\textwidth]{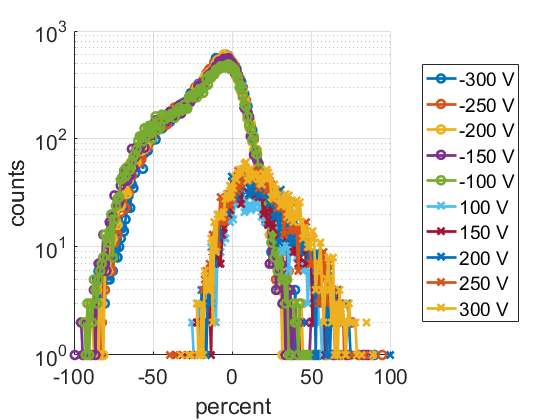}
	\caption{Data excluding split events.}
	\label{halo_b}
  \end{subfigure}
  
  \caption{Histogram of the summed charges collected in the 8 pixels surrounding a photon event divided by the energy collected in the central pixel. For both electrons and holes this halo is positive, i.e., it is of the opposite polarity of the signal in case of electron collection. }
  \label{halo}
\end{figure*}

In electron collection mode the determined gain increases slightly with increased sensor temperature. A much bigger influence is the applied bias voltage. The gain plateaus to about
95~ADU~/~59.5~keV~$\approx$~1.6~ADU/keV for biases more negative than approximately -200~V, as shown in figure \ref{gain}.

It is unclear if the change in gain with applied bias is due to a change in drift time and the
associated trapping time in conjunction with the intricacies of the small pixel effect \cite{spe, sim1}, or if this is caused by another mechanism. 


In hole collection mode the gain is almost independent of both bias and temperature. This effect may be compounded by our cluster finding algorithm which sums up the contributions of all pixels in the large clusters. The photopeak is un-observable in the individual pixel histograms (figure \ref{hist_b}) and can only be reconstructed when summing all pixels of a cluster. The Am-241 photopeak is not at the same position for both polarities. We expect them to be at different positions, as electrons and holes have different mobility-lifetime products. We note that the position of the peak at 75-80~ADU is much higher than anticipated from the data presented later in this report. We cannot exclude that the visible peak is a consequence of a selection bias, where only events in which the photon converts close to the readout electrode are successfully reconstructed, as those suffer the least from hole trapping and have the most favorable weighting field due to the small pixel effect.

Figure \ref{halo} shows histograms of charge collected in the 8 pixels surrounding a photon event divided by the energy collected in the central pixel. The distribution is consistent with effects expected from hole trapping. Trapping of charge carriers drifting away from the pixel electrode are expected to induce an opposite polarity halo around the center of the event, while charge carriers trapped while drifting towards the electrode are expected to induce an equal polarity halo with respect to the polarity of the collected species \cite{review}. Thus, in case of electron collection, the collected signal is negative and the halo is positive; in the case of hole collection both signal and halo are positive. For electron collection, approximately -2\% of the charge in the central pixel is found on average in each pixel of the halo. This is an effect of the electrode geometry and the trapping of charge carriers. We observe it to be independent of the bias voltage. Since the calculation of expected effect magnitude of the geometry and trapping is beyond the scope of this investigation, we cannot exclude that other effects, e.g., charge spreading and charge sharing, contribute to the observed results.

\subsection{Noise}

\begin{figure*}[tb!]
  \centering
  \begin{subfigure}[t]{0.45\textwidth}
	\includegraphics[width=\textwidth]{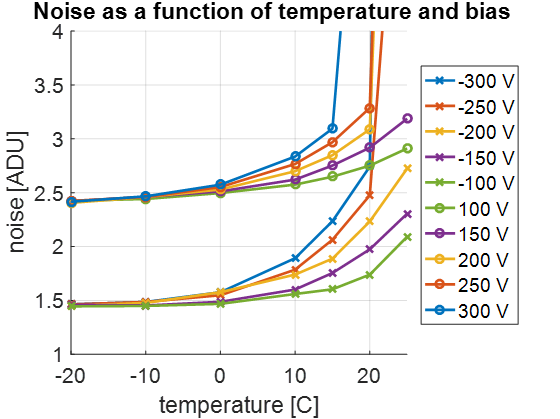}
	\caption{Average noise of the detector system as a function of temperature and bias.}
	\label{noise_a}
  \end{subfigure}
\quad
  \begin{subfigure}[t]{0.45\textwidth}
	\includegraphics[width=\textwidth]{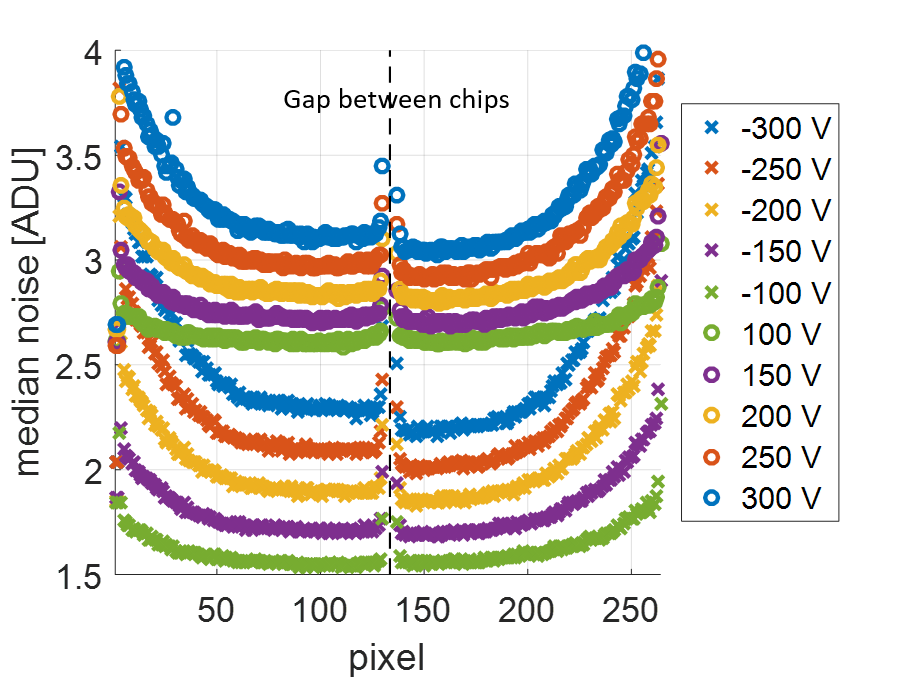}
	\caption{Median of the per pixel noise in each horizontal line of the detector as a function of bias and vertical pixel position at 15~C. }
	\label{noise_b}
  \end{subfigure}
  
  \caption{The noise figures as a function of position, temperature and bias.}
  \label{noise}
\end{figure*}

The noise of the system was determined by calculating the standard deviation from a series of dark images with an integration time of 100~$\upmu$s. The results are presented in figure \ref{noise}. Determining the noise this way accounts only for effects caused by the readout system and dark current. Additional noise caused by incomplete charge collection due to trapping or charge splitting between several pixels is not included.

As shown in figure \ref{noise_a}, the noise is a function of temperature and bias for temperatures above
10~C. This means that the dark current, and its associated shot noise, contribute significantly to the
overall noise at these temperatures. At lower temperatures the noise is dominated by the system’s
inherent readout noise, which is approximately equivalent in magnitude to a 1.2 keVx-ray in electron
collection mode.

Figure \ref{noise_b}, show that the noise is not distributed randomly in the detector, but that the noise profile clearly increases towards the sides of the detector. No significant influence of the horizontal pixel position is observed. Since the heat producing readout electronics are located towards the sides of the detector, it is likely that these components cause local heating effects, increasing the dark current and thereby the noise towards the edge of the detector. These effects have not been observed previously in MM-PAD systems as the dark current of both silicon sensors and Schottky type CdTe sensors is several orders of magnitude below the dark current in GaAs:Cr and therefore negligible compared to the system's read noise.

\subsection{Flat Field}


Previous studies of GaAs:Cr sensors using photon counting detectors found the detector response to be very non-uniform. Per pixel correction factors (i.e., flat field corrections) were used to compensate the non-uniformity of the sensor [7, 34, 35]. Here, all flat fields were determined after continuous illumination of the sensor for at least 10 seconds to exclude effects from incomplete settling of the sensor at low temperatures (see below for details). We define the correction factor, $c_{ff}(x, y)$, of the pixel with the coordinates $(x, y)$ in the following way:

\begin{equation}
c_{ff}(x,y) = \frac{ I(x,y)}{< I(mask) >} \label{c_ff}
\end{equation}

where $I(x,y)$ is the background subtracted measurement value of this pixel, and $< I(mask) >$ is the average background subtracted measurement value of the group of pixels in a reference region (mask). We choose the reference region to be the central region of chip 1 by excluding all pixels that are 5 pixels or less from an edge of the chip. 

\begin{figure*}[tb]
  \centering
  \includegraphics[width=1.0\textwidth]{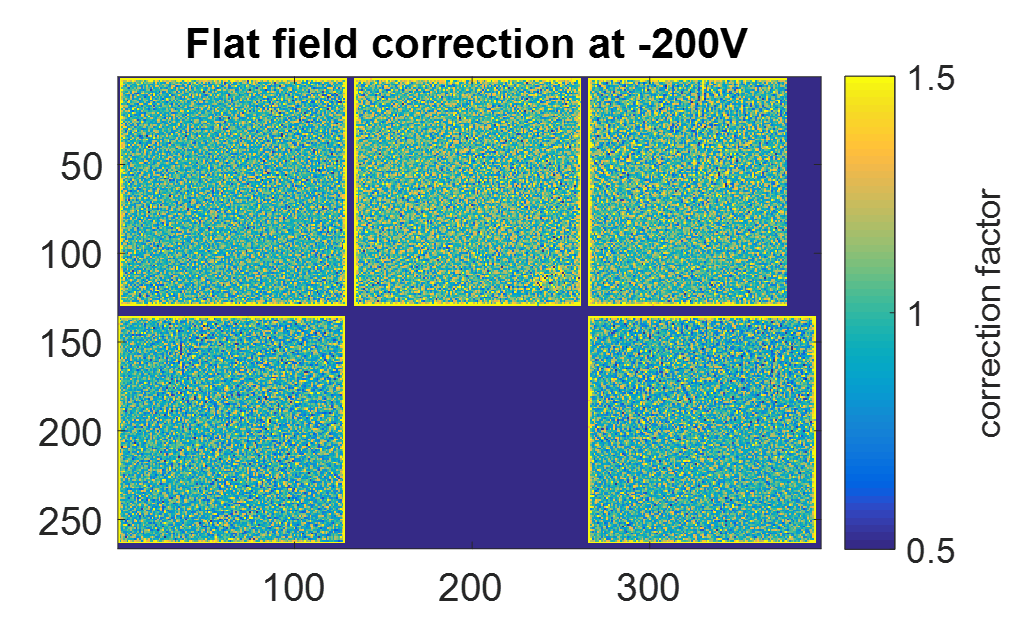}
  \caption{Typical flat field obtained at the preferred operating conditions of -200~V bias, 15~C sensor temperature and 47~kV tube voltage. The average intensity per pixel was about -150~ADU.}
  \label{ff_comp}
\end{figure*}

Choosing the reference region this way avoids the pitfall of normalizing each chip individually, thereby potentially losing information of relative shifts between the chips (about $\pm$~2\%), and avoids dealing with the imperfections of the other chips. Definition of the reference is inherently arbitrary as long as it excludes anomalous pixels and is big enough the provide a reasonable estimate of the mean. With proper masking and weighting information from all chips could have been used for this purpose without changing the results.

A typical map showing the flat field corrections for each pixel is shown in figure \ref{ff_comp}. Note that the spread of correction values is distributed within the range from 0.5 to 2, centered around the ideal value of 1, which is much larger than the spread of correction values observed in either CdTe (about $\pm$~20\% \cite{static}) or silicon (about $\pm$~2\% \cite{mm_gain}). 

\begin{figure*}[tb]
  \centering
  \includegraphics[width=0.8\textwidth]{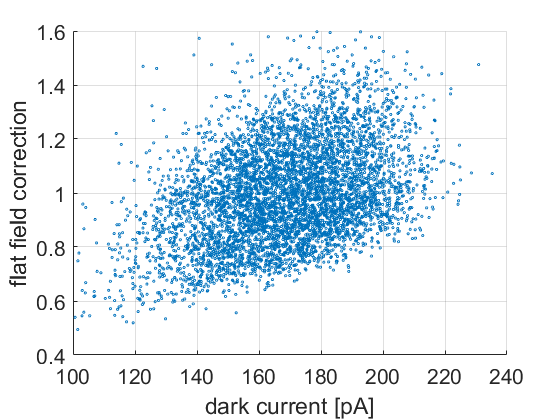}
  \caption{Correlation between dark current and flat field correction of chip 1 at the preferred operating conditions of -200~V bias, 15~C sensor temperature and 47~kV tube voltage. There is a weak (about 0.38) correlation between dark current and flat field correction.}
  \label{scatter}
\end{figure*}

The structure of the flat field map is not identical to the structure of the dark current map (figure \ref{dark}), although correlations between 0.35 (chip~3) and 0.49 (chip~6) exist (figure \ref{scatter} shows a scatter plot for chip~1). Some structures, like the line-like structure in the lower left corner of chip~3 and top-center of chip~5, seem to be present in both maps, but others, like the `grains' of high current do not have corresponding structures in the flat field correction map. This hints at the possibility that there may be more than one type of defect present in the material and that some of the defects responsible for the generation of the dark current may not solely be responsible for the observed structure in the flat field correction.

Since the flat field correction is dominated by effects of the effective pixel volume (see next section), we encourage the reader to think of the flat field correction not as a necessary correction of detector imperfections, but rather as a data transformation from absolute number of detected photons to photon density at a given position. This facilitates understanding, why the spectroscopic measurements presented before, determining gain and noise, should not be flat field corrected, while the results presented in later sections should. For many experimental applications application of a flat field correction will improve the data quality and can be done automatically, assuming the flat field correction factors are well determined for the experimental conditions (e.g. x-ray energy).

However, for other measurements, such as measuring intensities of spots of similar or smaller size than the characteristic size of the flat field variation, a flat field correction can increase the measurement uncertainty.

We determined flat field corrections for a wide range of experimental conditions in order to
investigate which factors contribute to changes in the flat field correction factors. To compare
different flat field distributions we used the following root-mean-square-deviation metric:

\begin{equation}
RMS(test) = \sqrt{\frac{1}{N_{pixel}}\sum_i \left(c_{ff,test}(i) - c_{ff,ref}(i)\right)^2} \label{rms}
\end{equation}

where a given `test' flat field, $c_{ff,tet}$, is compared to a `reference' flat field\footnote{Incidentally this happens to be at -200~V bias, 15~C and 47~kV tube voltage for all measurements, except the stability over time measurement. The reference flat field was reacquired for each measurement series.}, $c_{ff,ref}$, by calculating the quadratic mean of the per-pixel differences with $i$ being the pixel index. The lowest possible value of zero is obtained when comparing the reference field to itself, for all other comparisons the value is larger than zero, and a larger number indicates a larger deviation. Since the average over a larger region of the flat field itself is normalized to unity, the RMS deviation also indicates the relative deviation in flat field values.

\begin{figure*}[tb!]
  \centering
  \begin{subfigure}[t]{0.48\textwidth}
	\includegraphics[width=\textwidth]{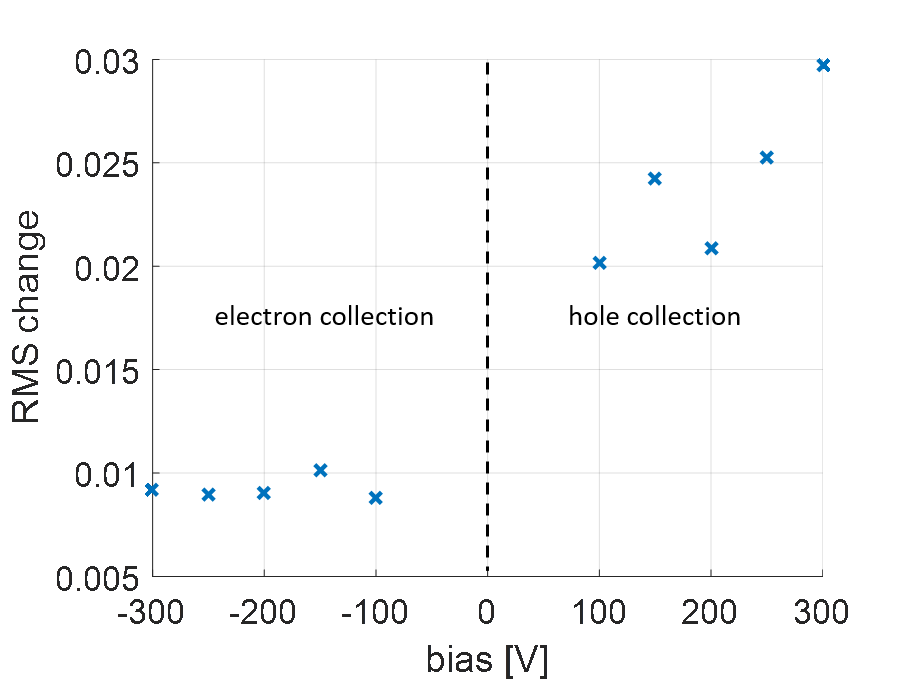}
	\caption{Comparison of flat field corrections taken after a week of testing including many bias cycles with pre-testing corrections. Comparing flat field corrections before and after temperature cycles produces similar results. The corrections are within the expected statistical spread for determining the values twice.}
	\label{ff_1_a}
  \end{subfigure}
\quad
  \begin{subfigure}[t]{0.48\textwidth}
	\includegraphics[width=\textwidth]{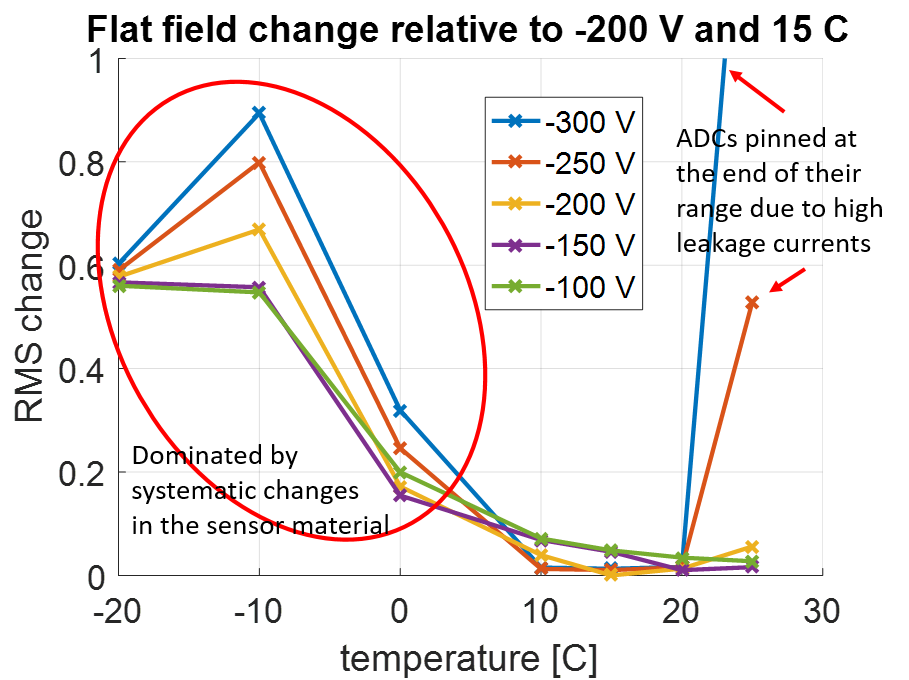}
 	\caption{RMS change in the flat field corrections as a function of temperature and bias for a fixed tube voltage of 47~kV. Within the range of 10~--~25~C flat field corrections are reasonably similar (RMS~<~0.1), at temperatures below 10~C changes become so significant that the metric loses meaning.}
	\label{ff_1_b}
  \end{subfigure}
  
  \caption{Changes of the flat field correction factors as a function of time, temperature and applied bias.}
  \label{ff_1}
\end{figure*}

Where possible we used 15~C temperature as a reference, as we found this temperature to be a compromise between higher temperatures, which generally seem to provide better uniformity of the response, and lower temperatures that allowed to sample a larger dynamic range.

\begin{figure*}[tb]
  \centering
  \includegraphics[width=1.0\textwidth]{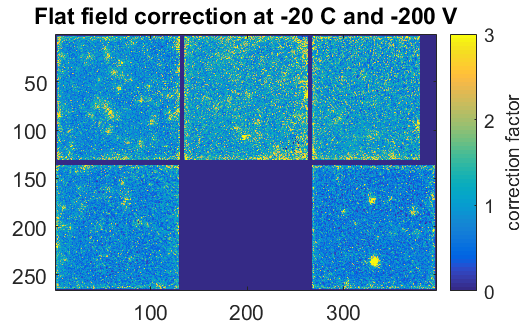}
  \caption{Typical flat field obtained at a bias of -200~V bias, -20~C sensor temperature and 47~kV tube voltage. The color scale has been adjusted to increase the contrast of the hotspots. Individual hotspots show flat field correction factors exceeding 12.}
  \label{ff_hs}
\end{figure*}

As shown in figure \ref{ff_1} the flat field correction is stable in time, but changes significantly with sensor temperature. Figure \ref{ff_1_a} shows the RMS change in the flat field after approximately one week of testing at a constant temperature of 15~C. The obtained RMS values are commensurate with the values expected due the statistical nature of the determination of the flat field correction (a few percent). However, changes in temperature show much more significant changes. As shown in figure \ref{ff_1_b} temperature changes in the range of 10~--~25 C show limited changes (RMS $<$ 0.1)\footnote{Except for high negative voltages at 25~C, which were pinned at the extreme of the available dynamic range.}. For temperatures below 10~C, hotspots start appearing making the flat field corrections are significantly different, as shown in figure \ref{ff_hs}, yielding very large RMS values. 

\begin{figure*}[tb]
  \centering
  \includegraphics[width=0.7\textwidth]{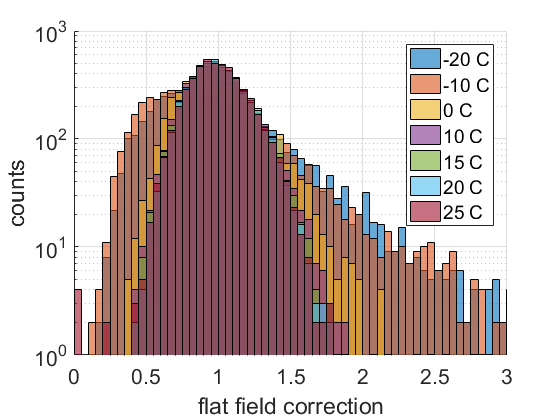}
  \caption{Distribution of the flat field correction factors at an applied bias of -200~V and a fixed tube voltage of 47~kV as a function of temperature. For temperatures below 10~C hotspots start to appear which skew the distribution significantly.}
  \label{ff_temp}
\end{figure*}

Analysis of the distribution of correction factors as a function of temperature (the histograms are shown in figure \ref{ff_temp}) shows that the hotspots appear gradually around 0~C and skew the entire distribution towards high correction factors. The data appears to be consistent with strong lateral fields that decrease in strength at higher temperatures. An electrically active defect that gets passivated due to trapping of charge carriers from the dark current in the material could be a possible cause, but the exact nature the hotspots remains unclear.

\begin{figure*}[tb!]
  \centering
  \begin{subfigure}[t]{0.45\textwidth}
	\includegraphics[width=\textwidth]{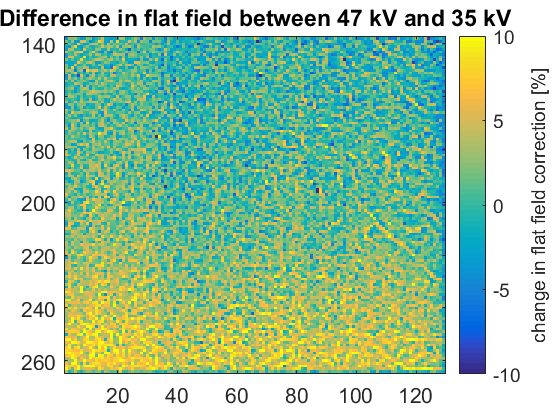}
	\caption{Absolute change in the flat field correction between 47~kV and 35~kV tube voltage for -200~V bias at a temperature of 15~C. To enhance the visibility of the resulting structures only chip 2 (lower left) is shown.}
	\label{ff_2_a}
  \end{subfigure}
\quad
  \begin{subfigure}[t]{0.45\textwidth}
	\includegraphics[width=\textwidth]{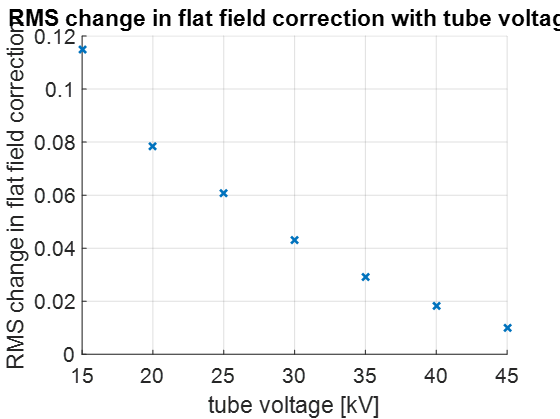}
	\caption{RMS change of flat field corrections relative to a set of flat field corrections determined at 47~kV tube voltage as a function of tube voltage for a constant bias of -200~V at a temperature of 15~C.  }
	\label{ff_2_b}
  \end{subfigure}
  
  \caption{Changes in flat field correction as a function of x-ray tube voltage. Results are likely influenced by the different absorption profiles for different tube voltages, effectively probing different detector depths.}
  \label{ff_2}
\end{figure*}


Previous reports using photon counting detectors show the flat field is sensitive to the energy of the incoming beam \cite{G6}. In order to establish the magnitude of this effect for our integrating detector system, the tube voltage (and thereby the average and maximum photon energy) was varied, while keeping the applied bias and temperature constant.

Figure \ref{ff_2_a} shows the relative change in the flat field when changing the tube voltage from 47~kV to 35~kV. The observed structure shows many long `lines' of increased or decreased response, which is a different structure from either the dark current or the flat field distributions. The lines do not correspond to structures in the ASIC and change in position from chip to chip, so an influence of the ASIC on the appearance of the lines is unlikely.

Figure \ref{ff_2_b} shows the RMS change in the flat field distribution as a function of tube voltage. Using a reference at 47~kV, -200~V and 15~C, we observe that the changes in the flat field increase with increasing difference between the `test' tube voltage and the reference tube voltage. This means that in order to properly correct measurement data using a flat field correction, the flat field should be determined at the energy of the experiment. While this approach is comparatively straightforward using monochromatic x-rays, using a polychromatic beam like from an x-ray tube incurs additional complications due to hardening of the beam caused by the sample. 

Keeping in mind that higher photon energies penetrate further into the sensor material, it is plausible that the observed effects are caused by the change in effective weighting of the contributions from different detector depths.

\subsection{Effective pixel size variations} 

\begin{figure*}[tb!]
  \centering
  \begin{subfigure}[t]{0.45\textwidth}
	\includegraphics[width=\textwidth]{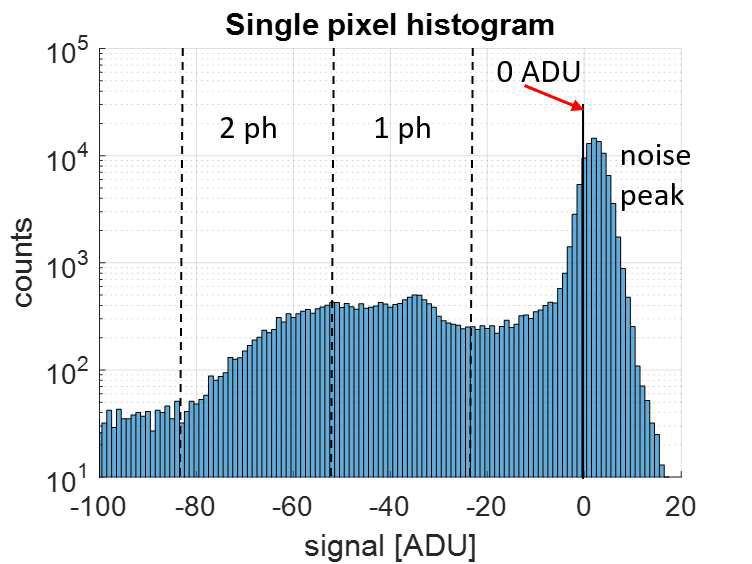}
	\caption{Histogram of the data from a single pixel for a single position step. The displayed data are from a pixel when the pinhole is centrally aligned. Counting thresholds are indicated to guide the eye. Note the noise peak is shifted towards positive ADU and the flat histogram profile at the thresholds, which makes the counting approach more robust against such offset shifts.}
	\label{ph_1_a}
  \end{subfigure}
\quad
  \begin{subfigure}[t]{0.45\textwidth}
	\includegraphics[width=\textwidth]{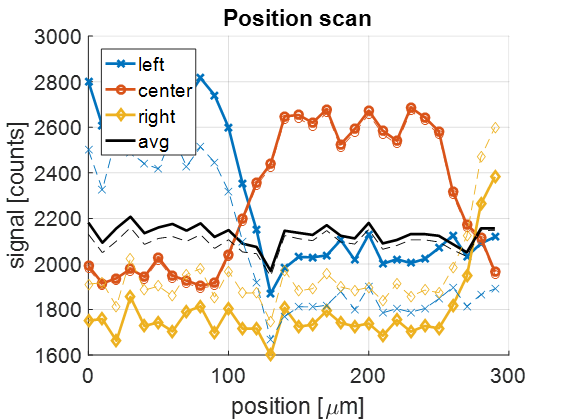}
	\caption{Photon counts in three adjacent pixels pixels for each scan position. Dashed lines indicate the results when the flat field correction is applied to the number of counts for each individual pixel. The average signal of all three pixels (black lines) does not show a significant change when scanning over the boundary between the pixels, indicating proper counting of split events. }
	\label{ph_1_b}
  \end{subfigure}
  
  \caption{Results from scanning a 25~$\upmu$m pinhole horizontally across the pixel matrix in steps of 10~$\upmu$m.}
  \label{ph_1}
\end{figure*}

As outlined in the previous sections, there are indications that the variations in the flat field correction factors are caused by different effective pixel sizes. This means the assumption that each pixel of the ASIC (150~$\upmu$m pitch in each direction) collects the charges generated within a perpendicular (box-shaped) volume defined by the ASIC pitch in x and y and the sensor thickness in z is no longer valid\footnote{Strictly speaking this assumption is never valid, as there are always complications like surface effects and diffusive exchange with neighboring pixels. Nevertheless it is a useful approximation in many cases.}. The collection volume is replaced by an effective volume that depends on temperature, bias and photon energy, as shown above.

To measure the effective pixel width, i.e., the effective horizontal pixel size, we used a pinhole array with 1~mm hole pitch that consisted of a 10-by-10 grid of 25~$\upmu$m diameter pinholes in a 75~$\upmu$m thick tungsten substrate. This pinhole mask was aligned with the pixel centers and then moved in front of the detector with 10~$\upmu$m steps along the horizontal pixel row direction. 

At 47~kV tube voltage the pinhole mask substrate is not completely opaque. Therefore the contrast between pinhole and substrate is very low and we used the thresholding technique \cite{th1, th2} to reduce noise and recover a usable signal from the data. Figure \ref{ph_1_a} shows the single pixel histogram and the chosen thresholds for a pixel under a pinhole. The thresholds were adjusted such that no under- or over-counting is observed when scanning over pixel boundaries (see figure \ref{ph_1_b}).

Note that the source produces a polychromatic spectrum that is further hardened by the transmission of the mask. Thus the signal, which is measured in counts, is a weighted measure of the total flux suppressing noise and counts do not necessarily correspond to individual photons. This technique may produce a different number of counts than the circuitry of photon counting detectors due to the way split events and photon pileup during the integration time is handled. The signal in all three pixels shown in figure \ref{ph_1_b} were corrected using a flat field obtained without taking the beam hardening of the mask into account and agree within approximately 5\% after the correction. A better agreement is expected if the hardening of the mask were taken into account.

Since the data are corrected for the signal induced offset (see below for details on this correction) using the entire substrate area as a reference, pixels underneath a pinhole show an additional uncompensated offset (shift of the noise peak to positive ADU) due to the additional signal (and hence photo current). The selected thresholds are in comparatively flat regions of the photon histogram, thus shifting the histogram by a few ADU either way does not appreciably change the overall signal.

\begin{figure*}[tb!]
  \centering
  \begin{subfigure}[t]{0.45\textwidth}
	\includegraphics[width=\textwidth]{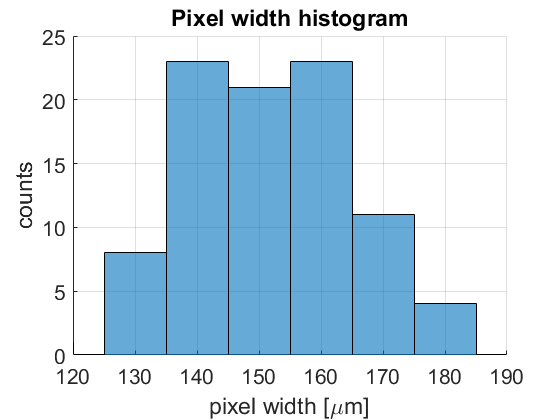}
	\caption{Effective pixel width determined from all pinholes where pixel width determination was possible. Although the nominal width of 150~$\upmu$m is still the most probable width, there is a significant spread in the effective pixel width.}
	\label{ph_2_a}
  \end{subfigure}
\quad
  \begin{subfigure}[t]{0.45\textwidth}
	\includegraphics[width=\textwidth]{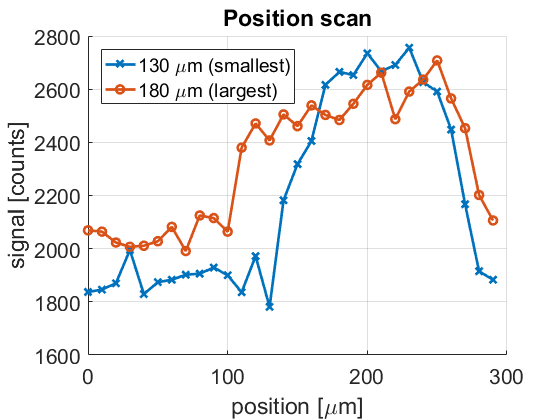}
	\caption{Scan profiles for two given pixels with the smallest and largest effective pixel width. The corresponding flat field correction factors are below and above unity for the large and small pixel, respectively.}
	\label{ph_2_b}
  \end{subfigure}
  
  \caption{Distribution of effective pixel width and examples of scan profiles for a small and a large pixel.}
  \label{ph_2}
\end{figure*}

We determine the effective width of the pixel by measuring the distance between the points where the additional signal due to the pinhole crosses on half of its value. Sufficient data were collected to reconstruct the width of 92\% of the pixels under the pinholes. A histogram of the distribution of reconstructed pixel sizes is shown in figure \ref{ph_2_a}. While the average size of 151~$\upmu$m corresponds to the nominal pixel size, an effective width of 150~$\upmu$m is not the most commonly observed size. The entire distribution shows a standard deviation of approximately $\pm$ 15~$\upmu$m. Figure \ref{ph_2_b} shows the scan results for one example pixel of the smallest and largest size.

Note that the presented sample of 92 pixels represents only approximately 0.5\% of the total matrix and are preferentially located in the central area of the array. We cannot exclude that this selection biases the observed effects in a systematic way.

\subsection{Edge spread function}

To gain insight into the spatial resolution of the system the Edge Spread Function (ESF) was measured using a 6~mil (about 150~$\upmu$m) thick tantalum knife edge misaligned by 3 degrees with respect to the pixel matrix. This misalignment allowed sampling of the edge in five rows simultaneously for a better result. 

\begin{figure*}[tb!]
  \centering
  \begin{subfigure}[t]{0.45\textwidth}
	\includegraphics[width=\textwidth]{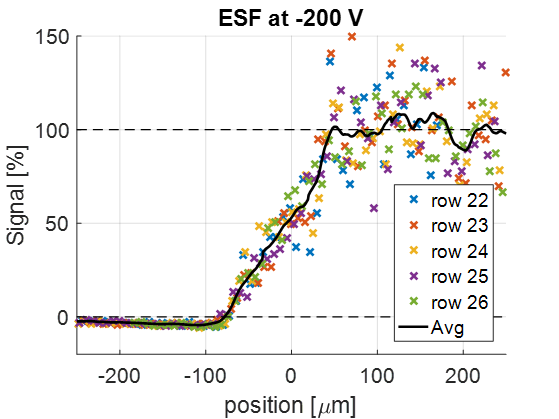}
	\caption{No correction.}
	\label{esf_1_a}
  \end{subfigure}
\quad
  \begin{subfigure}[t]{0.45\textwidth}
	\includegraphics[width=\textwidth]{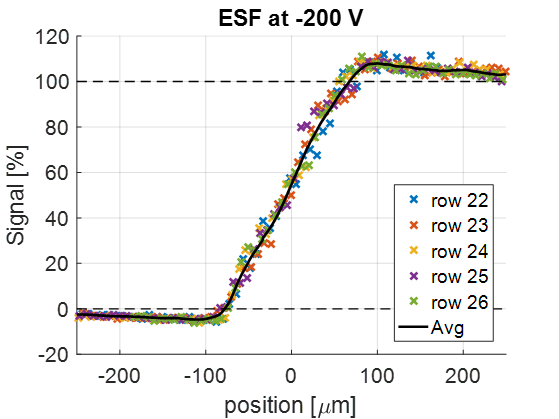}
	\caption{Pixel values are flat field corrected.}
	\label{esf_1_b}
  \end{subfigure}
  
  \caption{Comparison of the Edge Spread Function (ESF) at -200~V bias, 15~C and 47~kV tube voltage with and without flat field correction.}
  \label{esf_1}
\end{figure*}

Figure \ref{esf_1} shows the results for a fixed bias of -200~V, 47~kV tube voltage and 15~C operating temperature. Figure \ref{esf_1_a} shows the projection of all measurement values onto the relative distance of the knife edge from a pixel center. Figure \ref{esf_1_b} shows the same data, but after flat field correction. The solid black lines in both figures indicate the result after linearly interpolating the measurement values on a fine grid, shifting them to identical center values and averaging the data of all 5 curves. Flat field corrections were obtained under identical conditions (bias, tube voltage, temperature), but without the presence of the tantalum edge.

Since figure \ref{esf_1} clearly demonstrates the need for a flat field correction, all further results in this section are flat field corrected. All curves are normalized such that 100\% signal corresponds to the average value of the full signal far away from the edge. 100\% signal is about -150~ADU for electron collection and about 50 ADU for hole collection. Likewise 0\% signal corresponds to the signal under the shield\footnote{Which is close to zero, but not entirely, due to the small transmission of tantalum at high energy.} far away from the edge.

\begin{figure*}[tb]
  \centering
  \includegraphics[width=0.8\textwidth]{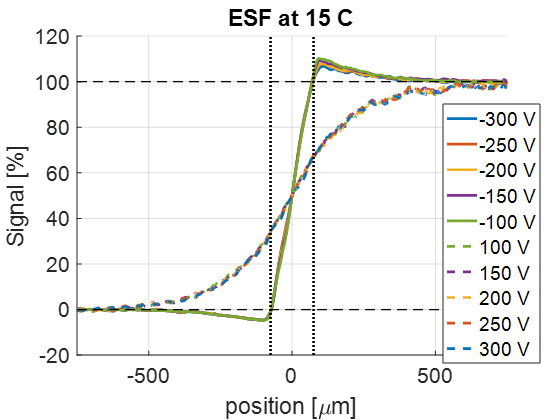}
  \caption{Edge Spread Function for 15~C and 47~kV tube voltage as a function of sensor bias. The magnitude of the over- and undershoot is comparable to the magnitude of the average halo determined from the Am-241 data. Vertical dotted lines indicate the pixel pitch for comparison.}
  \label{esf_3}
\end{figure*}

After flat field correction the ESF does not significantly depend on the magnitude of the applied bias\footnote{However, as shown above, the flat field correction depends on the applied bias.}. It does depend on polarity, as shown in figure \ref{esf_3}. For electron collection (negative polarity) we observe an overshoot in the tails of the ESF. For hole collection (positive polarity) we notice a very wide distribution on the order of 1~mm width. These results are consistent with the previous observation of a negative correlation between illuminated pixels and their neighbors (halo) for single events from an Am-241 source in case of electron collection and the observed large clusters in case of hole collection.

The results are also consistent with trapping and detrapping of charge carriers moving away from the electrode at negative bias (holes) and very low charge collection efficiencies for these carriers \cite{review}.

\begin{figure*}[tb!]
  \centering
  \begin{subfigure}[t]{0.45\textwidth}
	\includegraphics[width=\textwidth]{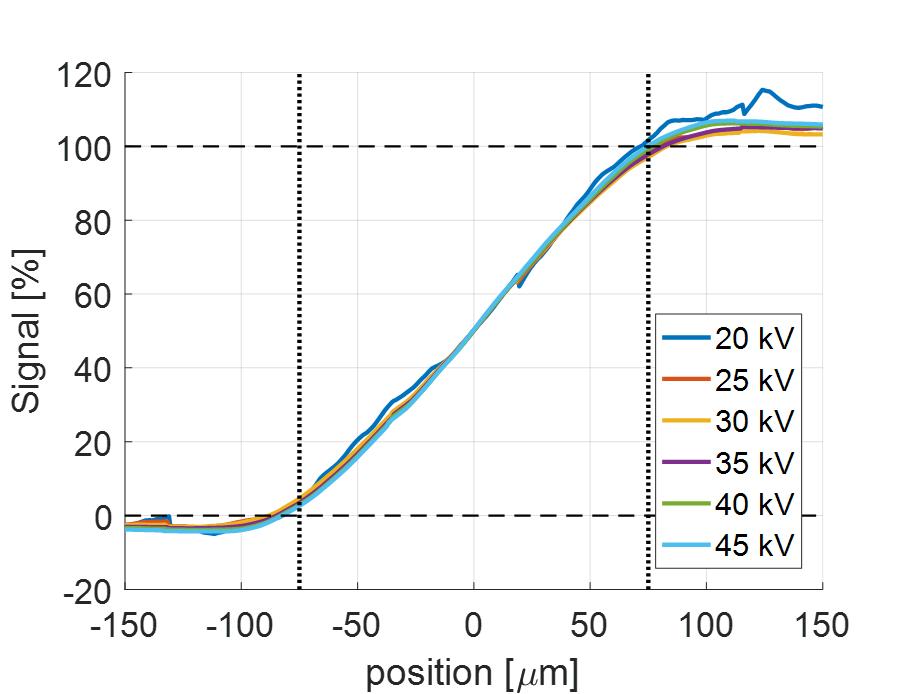}
	\caption{ESF as a function of tube voltage at a fixed temperature of 15~C.}
	\label{esf_2_a}
  \end{subfigure}
\quad
  \begin{subfigure}[t]{0.45\textwidth}
	\includegraphics[width=\textwidth]{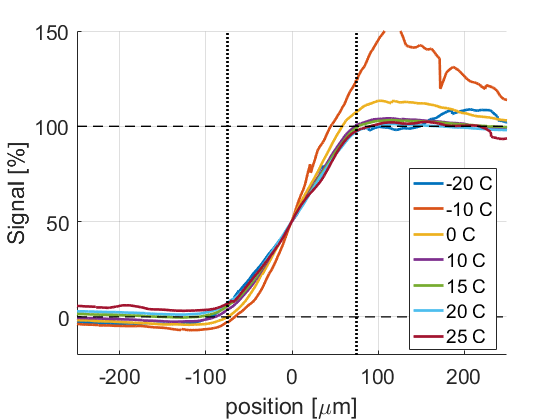}
	\caption{ESF as a function of temperature at a fixed tube voltage of 47~kV.}
	\label{esf_2_b}
  \end{subfigure}
  \caption{Edge Spread Function for a fixed bias of -200~V as a function of sensor bias. Vertical dotted lines indicate the pixel pitch for comparison.}
  \label{esf_2}
\end{figure*}

After correction, the ESF is also not a significant function of the tube voltage, as shown in figure \ref{esf_2_a} for a fixed bias of -200~V at a constant temperature of 15~C. Likewise for temperatures above approximately 10~C the ESF remains almost constant. However, for lower temperatures the ESF becomes more complex, as shown in figure \ref{esf_2_b}. The shown ESFs at low temperature are most likely no longer representative for the entire area due to the appearance of the hotspots.

\subsection{Rate dependent offset shift}

\begin{figure*}[tb!]
  \centering
  \begin{subfigure}[t]{0.45\textwidth}
	\includegraphics[width=\textwidth]{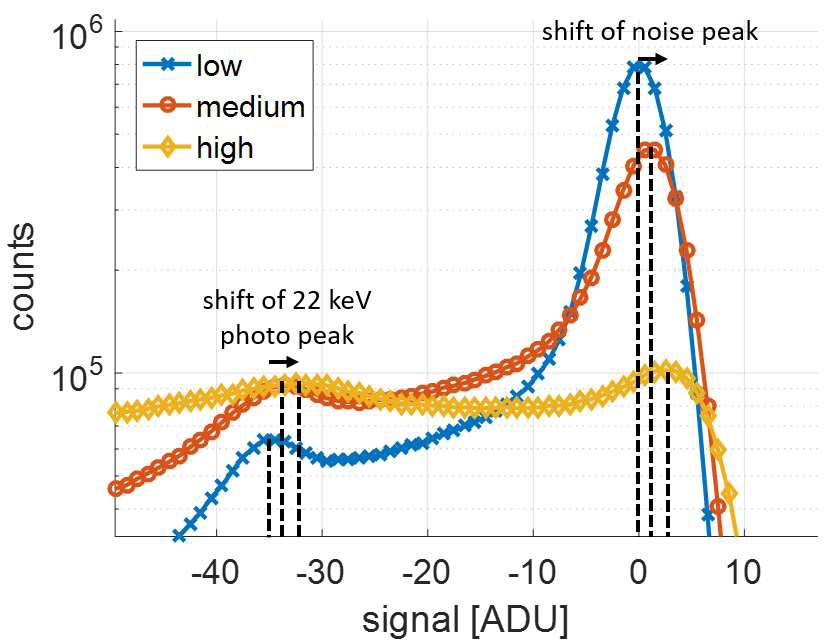}
	\caption{Photon histograms for three different low flux illuminations showing the shift in the histograms for different illumination intensities.}
	\label{shift_a}
  \end{subfigure}
\quad
  \begin{subfigure}[t]{0.45\textwidth}
	\includegraphics[width=\textwidth]{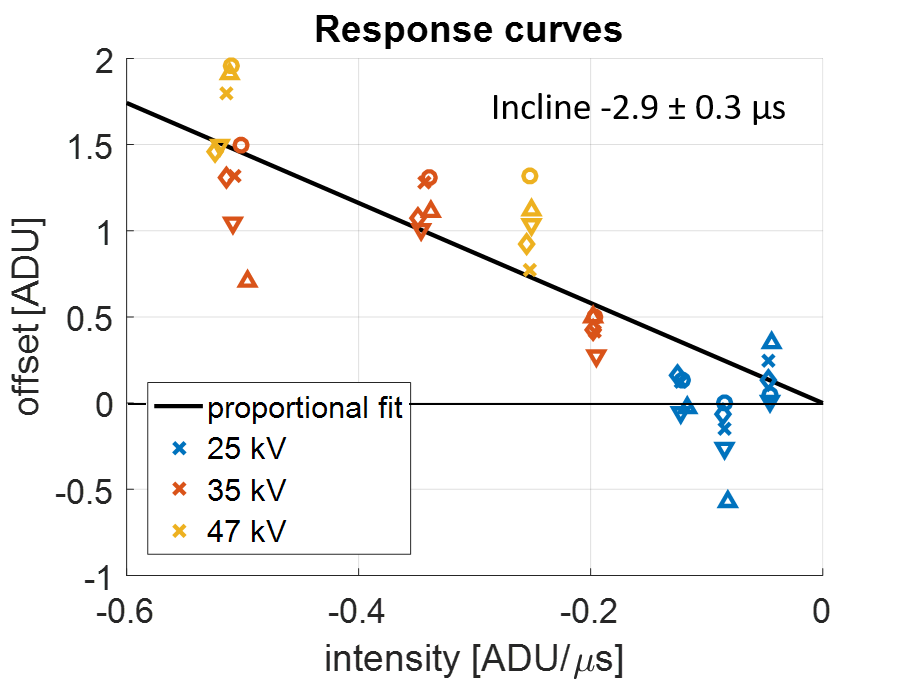}
	\caption{Offset shift as a function of intensity (ADU/$\upmu$s) for three different tube voltages. Each symbol denotes one of the 5 chips.}
	\label{shift_b}
  \end{subfigure}
  
  \caption{Example of the rate dependent offset shift at a fixed bias voltage of -200~V and a temperature of 15~C. To observe this shift the data were not common mode corrected.}
  \label{shift}
\end{figure*}

As indicated above and seen previously by others \cite{lpd}, there is an additional offset shift that is proportional to the intensity\footnote{Here we understand intensity as the ability of a given photon beam to cause photocurrent. It is proportional to the number of photons per unit area and unit time and proportional to the (average) photon energy.} of the incoming beam. This additional shift is likely caused by trapping charge carriers from the average photocurrent and thus cannot be corrected by the standard dark current subtraction.

An example of this is shown in figure \ref{shift_a}, where histograms of low flux data for three different tube currents are displayed. Except for the tube current, which was 0.2, 0.4 or 0.8~mA, all other conditions are identical. Looking at the position of the noise peak we observe that it moves towards higher ADU for higher intensities, i.e. the opposite direction of the signal, indicating that a net positive charge is induced on the readout electrode.

Figure \ref{shift_b} shows the magnitude of this offset shift for a fixed bias of -200~V and a temperature of 15~C. The data are consistent with a proportional response as a function of intensity in units of ADU/$\upmu$s\footnote{This is a measure of the photocurrent.}. The proportionality constant was determined to be -2.9~$\pm$~0.3~$\upmu$s, independent of the tube voltage.

Note that under the considered equilibrium conditions an average photocurrent is present, even though individual events may be without photons. Thus the offset also effects the position of the noise peak (absence of photons). This observation implies that the cause of the shift changes on a timescale that is large compared to the integration times investigated here (up to 100~$\upmu$s).
It is reasonable to assume that the detrapping times are large compared to the integration time (otherwise we would not be able to observe trapping effects). Therefore the additional offset may be caused by trapped charges that induce a non-zero signal. As shown below, the offset changes for non-equilibrium illumination in a complicated fashion, possibly indicating contributions to this effect by the trapping of both types of charge carriers.

\section{Characterization under non-equilibrium conditions}
The previous section presented results for equilibrium conditions. It is also important to understand the response to non-equilibrium conditions. Especially when dynamic processes are under investigation it is important that the response of the sensor material is reproducible and preferably identical to results obtained under equilibrium conditions. 

We created rapidly varying illumination conditions by interposing a moderately fast shutter in between the Ag tube and the detector. The opening and closing time of the mechanical shutter is about 15~ms. However, the time resolution of the measurement is set by the transit time of the
shutter past a single pixel, which is much less than 1~ms.

\subsection{Top hat illumination}

\begin{figure*}[tb!]
  \centering
  \begin{subfigure}[t]{0.45\textwidth}
	\includegraphics[width=\textwidth]{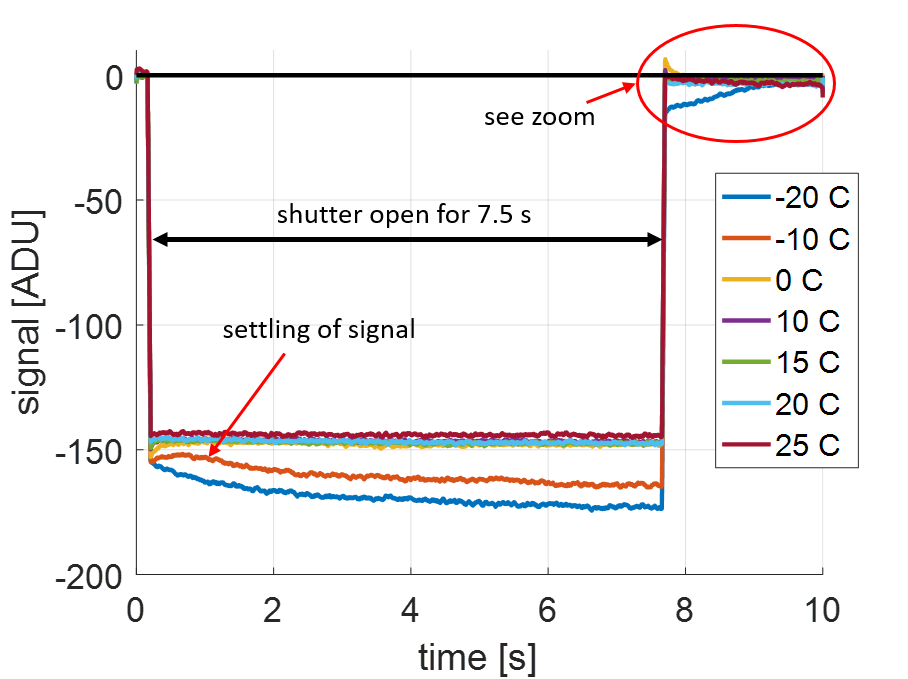}
	\caption{Whole recorded time series. For T~$<$~10~C significant settling times are observed.}
	\label{th_a}
  \end{subfigure}
\quad
  \begin{subfigure}[t]{0.45\textwidth}
	\includegraphics[width=\textwidth]{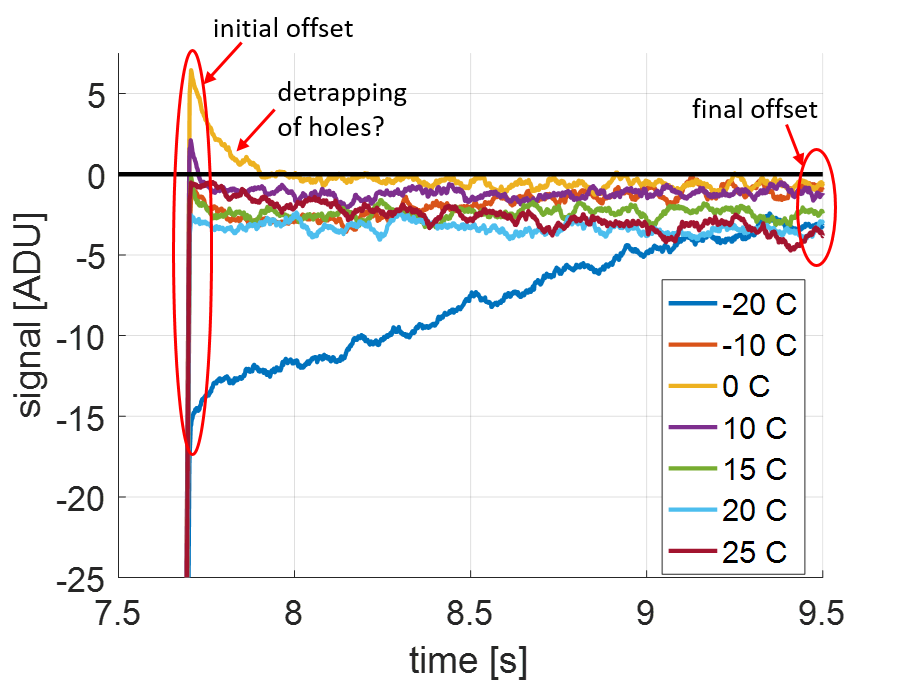}
	\caption{Zoom of the recorded data to show the complex decay structure after closing the shutter.}
	\label{th_b}
  \end{subfigure}
  
  \caption{Results for top hat illumination as a function of temperature at a fixed bias of -200~V and  tube voltage of 47~kV. The displayed data are not common mode corrected to show the dynamics of the offset change, but were smoothed with a 45~ms moving average filter to reduce the frame to frame noise.}
  \label{th}
\end{figure*}

We set up the detector to record a frame with 100~$\upmu$s integration time every 1~ms and took a total of 10,000 consecutive frames, corresponding to a total recorded duration of 10~s. During this time the sensor was illuminated for 7.5 seconds with 47~kV tube voltage and an intensity corresponding to approximately -150~ADU per frame. Figure \ref{th_a} shows results for the average response of the sensor as a function of sensor temperature during the recorded time.

We observe that the sensor has a significant settling time and complex decay behavior for temperatures below 10~C. This change in response for low temperature is consistent with the observed major changes in the flat field and the ESF at low temperature due to the appearance of hotspots.

Closer inspection of the settling after the closing of the shutter (see figure \ref{th_b}) shows that the signal exhibits an overshot of positive polarity. Furthermore the signal does not settle to zero but to a slightly negative value. Note that the common mode correction (see above for details) had to be turned off to reveal these dynamics, as they happen in all (illuminated) pixels at the same time.

\subsection{Decaying offsets}

\begin{figure*}[tb!]
  \centering
  \begin{subfigure}[t]{0.45\textwidth}
	\includegraphics[width=\textwidth]{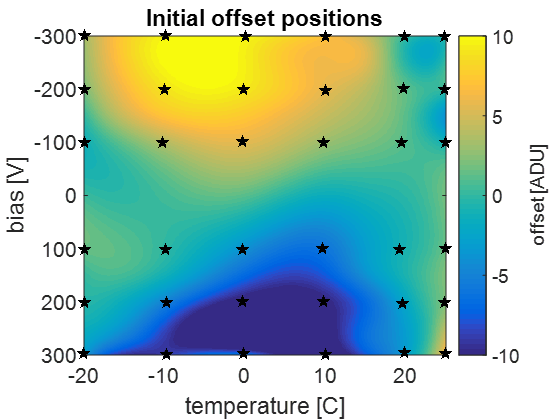}
	\caption{Initial offset.}
	\label{decay_map_a}
  \end{subfigure}
\quad
  \begin{subfigure}[t]{0.45\textwidth}
	\includegraphics[width=\textwidth]{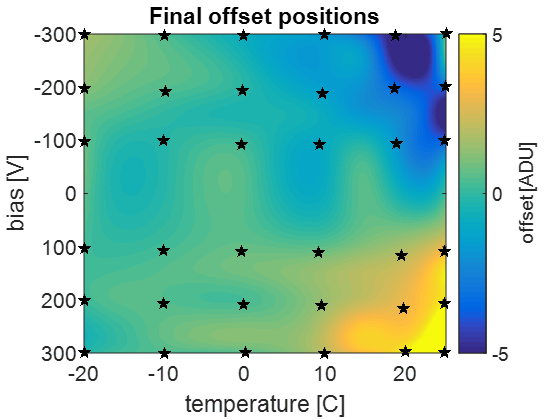}
	\caption{Final offset.}
	\label{decay_map_b}
  \end{subfigure}
  
  \caption{The offset after closing the shutter can be separated into an initial offset, determined right after closing the shutter and a `final' offset determined at the end of the recorded time frame. The shown data points (black stars) were obtained for a tube voltage of 47~kV and interpolated onto the shown maps.}
  \label{decay_map}
\end{figure*}

Further change is revealed by tracking the offset after illumination for an extended period of about 2 seconds. These changes are possibly caused by the detrapping mechanics in the material. Here we distinguish between the offset right after the shutter closes (initial offset) and the offset towards the end of the observed period (final offset). The initial offsets corresponds to the rate induced offset described above.

%

The change from initial to final offset is a strong function of temperature and bias, as shown in figure \ref{decay_map}. The data shown in figure \ref{decay_map_a} does not track the data displayed in figure \ref{th_b} exactly, as these date were smoothed for the plot, thereby reducing the initial peak. 

Re-taking these data with optimized parameters (temperature, total illumination time, integration time and total observation time) would possibly establish how long the detector needs to recover from the illumination and possibly allow one to track the detrapping of both electrons and holes and hopefully provide insight in detrapping times and into the involved trap energies. But this is beyond the scope of this paper.

\section{Imaging examples}

As many of the previously mentioned effects are difficult to visualize we provide a few examples of their practical implications in this section.

\subsection{Resistivity map}

\begin{figure*}[tb!]
  \centering
	\includegraphics[width=\textwidth]{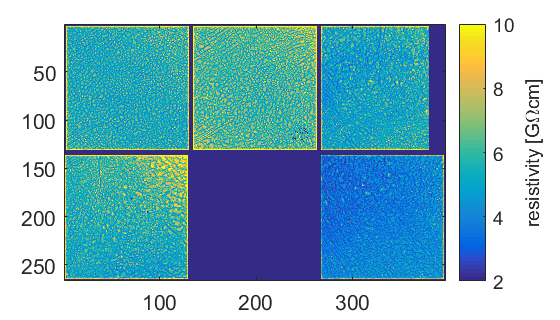}
  \caption{Resistivity map at 15~C. The resistivity was determined from integration time scans in absence of illumination at an applied voltage of -200~V. The average resistivity at this temperature is 5.4~G$\Omega$cm $\pm$ 30\%}
  \label{res}
\end{figure*}

As shown in the beginning of this paper, we can extract the dark current per pixel from taking dark images and determining slope of the offset by varying the integration time. The slope of the offset is proportional to the dark current.

Further, we can extract the resistivity per pixel, $\rho$, from the dark current per pixel, $I$, by normalizing to the average pixel size (using the flat field correction $c_{ff}$) and accounting for bias voltage, $U$, nominal pixel area, $A$, and sensor thickness, $d$ in the following way:

\begin{equation}
\rho = \frac{U A c_{ff}}{I d} \label{rho}
\end{equation}

The results are presented in figure \ref{res}. Since the dark current and flat field correction were determined at 15~C the average resistivity of 5.4~G$\Omega$cm $\pm$ 30\% is larger than the average resistivity of 0.5~G$\Omega$cm determined by the manufacturer (TSU) at room temperature.
The resistivity map shows many grain-like hot and cold spots that are not present in the flat field correction.


Assuming that the thermal current is produced by a mechanism similar to the one observed in
radiation damaged silicon sensors, namely deep traps close to the middle of the band gap, we expect
a temperature dependence that is proportional to  $T^2 exp(-E_{gap}/2kT)$ \cite{ac}. Thus, when cooling
from room temperature to 15 C we expect the current to decrease by a factor of approximately 3.
The observed increase in resistivity by a factor of approximately 10 is much larger than the expected increase using the mentioned model. Therefore it is unlikely that the observed dark current is generated by defects deep in the band gap.


\subsection{Line pair mask}

\begin{figure*}[tb]
  \centering
  \includegraphics[width=1.0\textwidth]{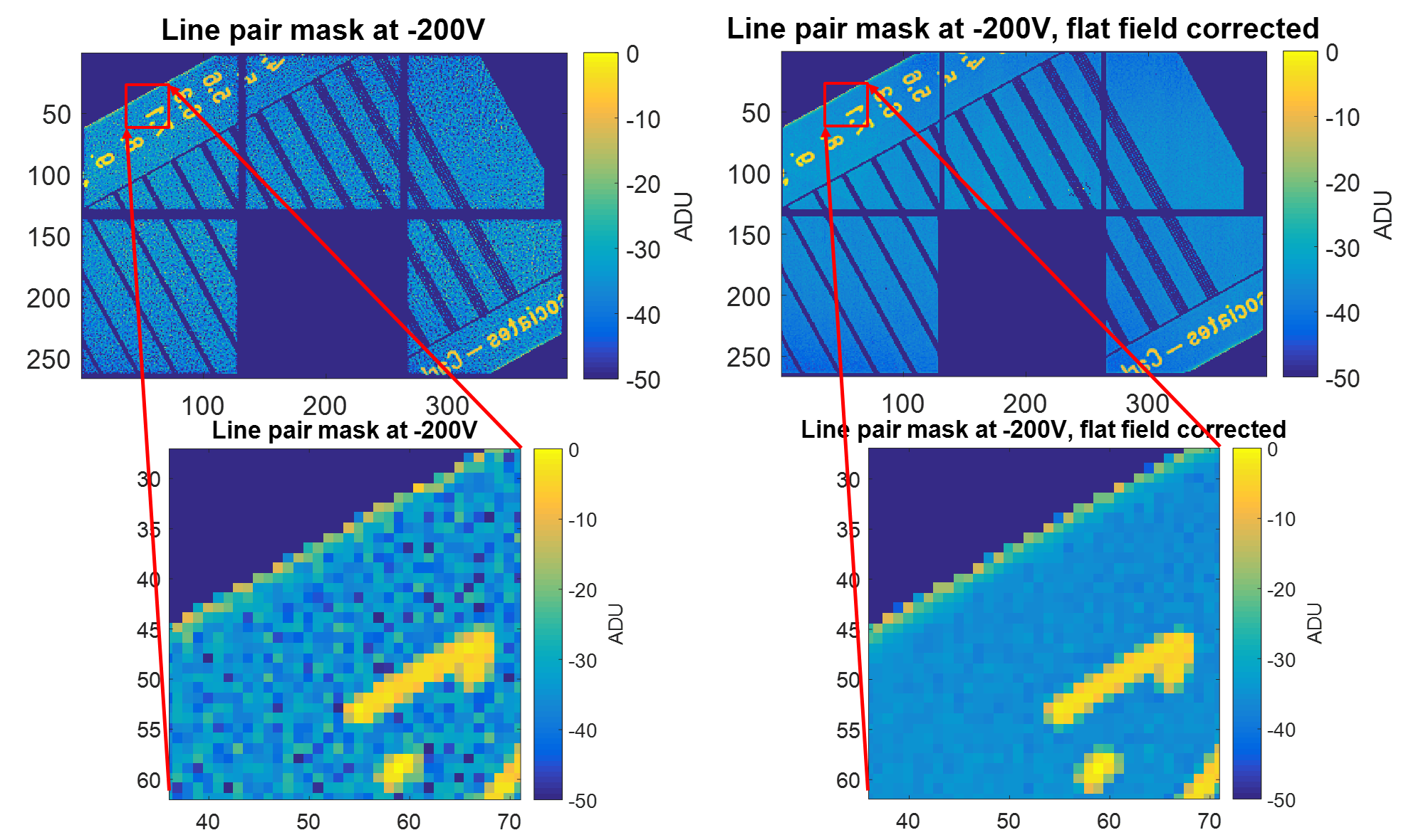}
  \caption{Images of a line pair mask with (right) and without (left) flat field correction. Inserts show the same zoomed-in area of both images. Data were acquired at -200~V bias, 15~C temperature and 47~kV tube voltage. The images are an average of 10,000 individual frames to reduce the overall noise of the system.}
  \label{lp}
\end{figure*}

Figure \ref{lp} shows an image of a standard line pair mask with and without flat field correction. Zoomed in regions are also displayed to further highlight the apparent `noise' of the response.

In the zoomed insert the `edge-enhancement' effect is visible on the edge of the mask. The image appears to indicate a thicker edge compared to the main area of the mask, which is not present in the real mask. This effect is caused by the `overshoot' observed in the ESF measurements. The more intense area beyond the mask creates an opposite polarity tail at the boundary with the mask that effectively cancels out some of the signal in these pixels.

\subsection{Watch}

\begin{figure*}[tb]
  \centering
  \includegraphics[width=1.0\textwidth]{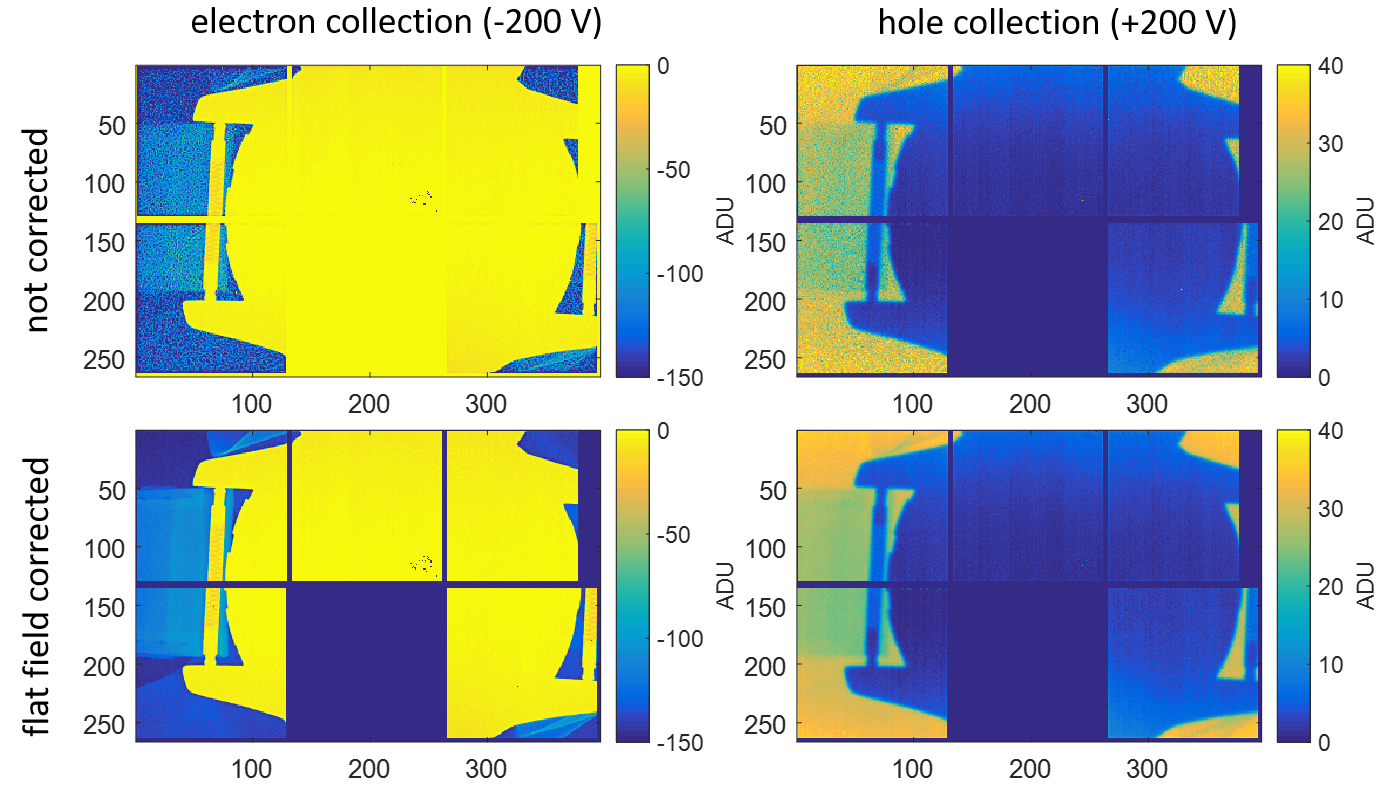}
  \caption{Radiographs of a watch with (lower row) and without (upper row) flat field correction. The left column shows images using electron collection, the right column shows results for hole collection. Data were acquired at 15~C temperature and 47~kV tube voltage. The images are an average of 10,000 individual frames to reduce the overall noise of the system.}
  \label{watch}
\end{figure*}

\begin{figure*}[tb]
  \centering
  \includegraphics[width=0.9\textwidth]{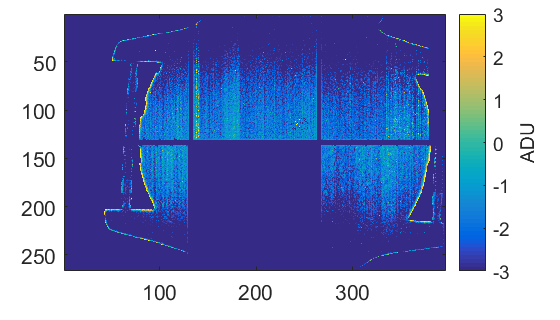}
  \caption{Flat field corrected radiograph of a watch with a narrow color scale around zero. The `edge enhancement' effect produces positive signal edges around the outline of the watch. Data were acquired with -200~V bias at 15~C temperature and 47~kV tube voltage. The image is an average of 10,000 individual frames to reduce the overall noise of the system.}
  \label{watch_2}
\end{figure*}

Radiographs of a wrist watch are displayed in figure \ref{watch}. The left column of images shows images collected in electron collection mode at -200~V bias, the right column shows the same situation in hole collection mode at +200~V bias. The top row shows uncorrected images, and lower row shows images after flat field correction. The corrected images are much clearer and additional details, such as the tape holding the watch in place become visible. Note that the color scale in electron collection covers 3.75 times the range of the color scale in hole collection mode. The lack of contrast in the images is simply because of the small dynamic range of the images.

Figure \ref{watch_2} shows the effect of the opposite polarity edge enhancement in electron collection mode (different color scale). The outline of the watch is clearly visible.

%

\section{Conclusions}

We demonstrated the successful mounting of chromium compensated GaAs sensors onto MM-PAD ASICs. Although the ASIC was not designed for electron collection, we were able to run a series of characterization measurements in order to evaluate the usefulness of GaAs:Cr as a sensor material for integrating detectors.

Since GaAs:Cr sensors are a made from a defect rich material and built as photoresistors rather than photo diodes, we have observed dark currents that are several orders of magnitude larger than what is commonly seen in reverse biased silicon photodiodes. This presented the challenge of finding an operating point that balanced several competing effects. We chose operation at -200~V bias at a temperature of 15~C as a compromise.

The dark current of the sensor is a strong function of both temperature and applied bias voltage. Higher temperature and higher absolute voltage increase the dark current. Since increased dark current increases the system read noise and reduces the available dynamic range in our system, a low temperature and absolute bias are preferred.    

We observe that higher magnitude biases produce higher gain. This effect shows a saturation type behavior for electron collection (meaning there are diminishing returns for higher magnitude bias) and is almost negligibly small in the hole collection case.

We find that the flat field distribution is very broad and spreads from values of 0.5 to values of 2 or higher around the ideal value of 1. The flat field corrections for each individual pixel are likely correlated with the effective pixel size (i.e., the effective sensitive volume for each pixel) and depend strongly on temperature, bias and photon energy. We also observe that the corrections are stable in time and for many temperature and bias cycles.

The spatial response of the sensor (once it is flat field corrected) is dominated in a way that is consistent with the trapping and detrapping of holes. This results in specific image distortion signatures depending on the collected species of charge carrier. In electron collection mode a point like illumination is surrounded by an opposite polarity halo. In hole collection mode a point like illumination will be significantly blurred.

The strong trapping of the material also induces an additional rate dependent offset shift. This shift seems to be caused by the average photocurrent in each pixel and, contrary to the author's intuition, does not seem to depend on photon energy.

When exposed to changing illumination conditions, we observe settling of the signal at temperatures below 10~C. After closing of the shutter we observe that the signal induced offset shift has a complicated decay behavior depending on temperature and applied bias.

In general we notice a different behavior of the material at 0~C and below than at 10~C and above. We speculate that this change in behavior is related to a change in one or more of the defects in GaAs:Cr. We encourage the community to follow up on these observations in the hope of understanding the material better once the underlying trap mechanisms are known. 

A more thorough study of the spatial response as a function of incident energy would help to understand the three dimensional structure of the material. Our measurements indicate that it is overly simplistic to assume uniform response throughout the entire depth of the GaAs:Cr material.

Finally, the results of investigating a top-hat illumination function show that the material response to pulsed intense radiation bursts might differ from that to quasi-continuous illumination. This is important not only for operation at Free-Electron Laser sources, but also for synchrotons with an appreciable time gap in between pulses, e.g., when operated with few electron bunches of high charge. 

In summary: The studied chromium compensated GaAs material shows several features that need to be controlled in order to obtain a reliable response in integrating imaging detectors. This mandates good temperature uniformity in the sensor and very good temperature stabilization, as well as a carefully selected bias voltage. 

Once the proper operating conditions are met and the required corrections are applied, GaAs:Cr sensors have some advantages over CdTe sensors in the energy range of roughly 10~--~50~keV.  Namely that we did not see any evidence of polarization in our measurements and the absence of sensor generated fluorescence around 30~keV. On the other hand, it is clear that GaAs:Cr exhibits its own complex behavior. 

\acknowledgments


The authors would like to thank Anton Tyazhev and the entire team for the functional electronics
laboratory from Tomsk State University in Russia for producing the sensors and providing basic
measurements of their bulk properties, as well as valuable discussions and input. This research
is based on research conducted at the Cornell High Energy Synchrotron Source (CHESS), which
is supported by the U.S. National Science Foundation via NSF award DMR-1332208 and by
Department of Energy award DE-SC0016035 for x-ray detector research to S.M.G.. The MM-PAD
concept was developed collaboratively by the Cornell detector group and the former Area Detector
Systems Corporation of Poway, CA, USA.


\nocite{*}

\end{document}